# Lagrange statistics in systems/markets with price constraints:
# Analysis of property, car sales, marriage and job markets by Boltzmann distribution and Cobb Douglas function

**J. Mimkes, Th. Fründ, G. Willis\*,**

Physics Department, University of Paderborn, 33096 Paderborn, Germany

\*South Elmsall, W. Yorkshire, U.K.

**Abstract**
In stochastic systems (markets) with constraints (prices) all interactions may be derived from the Lagrange principle: $L = E + \lambda \ln w \rightarrow$ maximum! Data for the "markets" property, automobile sales, marriage and jobs were analyzed by two models: the Cobb Douglas production function and the entropy function $\ln w$. All results are in favor of entropy. As a result wealth, car sales, marriage or jobs are not a matter of reason but of chances. Similar to other disciplines (physics, chemistry, meteorology) the Lagrange principle may be regarded as basic equation of economics.

**Introduction**
Since the work of Pareto (1) as long ago as 1897, it has been known that economic distributions strictly follow power law decays. These distributions have been observed across a wide variety of data. So far different models like Walrasian and Marshallian theories of market equilibrium have been presented. Recently, Georgescu-Roegen (2), Foley (3), Weidlich (4) Mimkes (5,6), Levy and Solomon (7), Solomon and Richmond (8), and Kreft (9) have proposed statistical models to economic distributions. This paper goes back to the Lagrange principle of statistics with constraints.

**1  Cobb Douglas production function**
The Lagrange equation

$$\lambda\, f\; +\; g \quad \rightarrow \quad \text{maximum!} \tag{1.0}$$

is widely accepted as basic equation of economics: it applies to all functions (f) that are to be maximized under constraints g. The distribution of wealth, the employment of people, the production of goods, economic actions (f) have to be optimized under the constraints of capital, costs or prices (g).
Eq.(1.1) is often given as

$$\lambda Y + \Sigma N_j k_j \rightarrow \text{maximum!} \quad (1.1).$$

Y is the production function that will be maximized under the constraints of the total capital $K = \Sigma N_j k_j$. $N_j$ is the number of people in the property class ($k_j$). $\lambda$ is the Lagrange factor and may be regarded as a mean capital level per person or standard of living of the economic system.

In many calculations a special Cobb Douglas production function Y is applied,

$$Y = A N_1^{\alpha_1} N_2^{\alpha_2} .. = A \prod N_j^{\alpha_j} \quad (1.2).$$

A is a constant, the exponents $\alpha_j$ are the elasticity constants with $\Sigma \alpha_j = 1$.
At equilibrium (maximum) the derivative of equation (1.1) with respect to $N_j$ will be zero, this leads to

$$\lambda \, \partial Y / \partial N_j - k_j = \alpha_j Y / N_j - k_j = 0$$

Solving this equation for $N_j$ we obtain

$$N_j(k_j) = \alpha_j Y / (k_j / \lambda) \quad (1.3).$$

The number $N_j$ of people in a property class (k j) decreases with rising property (k j). The total amount of capital $K_j$ in tax class (k j)

$$K_j(k_j) = k_j N_j(k_j) = \alpha_j \lambda Y = \text{constant} \quad (1.4)$$

depends only on the elasticity constant. The sum of all wages costs is given by

$$\Sigma K_j(k_j) = \Sigma N_j k_j = (\Sigma \alpha_j) \lambda Y = \lambda Y.$$

We may test the Cobb Douglas function (1.2) by applying equations (1.3 and 2.4) to the property data in Germany in 1993 in table 1.

**Distribution of property in Germany 1993 (DIW)**
Property data in Germany (1993) have been published (8) by the German Institute of Economics (DIW). The data show the number N(x) of households and the amount of capital K(x) in each property class (k), table 1.

| Property distribution in Germany 1993 (DIW estimated) |||
|---|---|---|
| Total property or capital | 9920 | Bill. DM |
| Number (N) of households | 35,6 | Mill. |
| Mean property ($\lambda$) per household | 278 | kDM / Hh |

| Property class (k) in kDM | Number N(x) of Hh in % | Property K(x) of Hh in % |
|---|---|---|
| =0 | 1,5 | 0 |
| $k_1$:   <100 | 44,5 | 9,5 |
| $k_2$:   100-250 | 24,7 | 17,6 |
| $k_3$:   250-500 | 20,3 | 28,1 |
| $k_4$:   500-1000 | 6,3 | 16,8 |
| $k_5$:   >1000 | 2,7 | 28 |

***Table. 1*** *Distribution of property in different property classes ($k_j$) for households in Germany (1993), according to study by the DIW (11).*

The data are generally presented by a Lorenz distribution. Fig. 1.1 shows the distribution of total capital K vs. the total number N of households according to table 1.

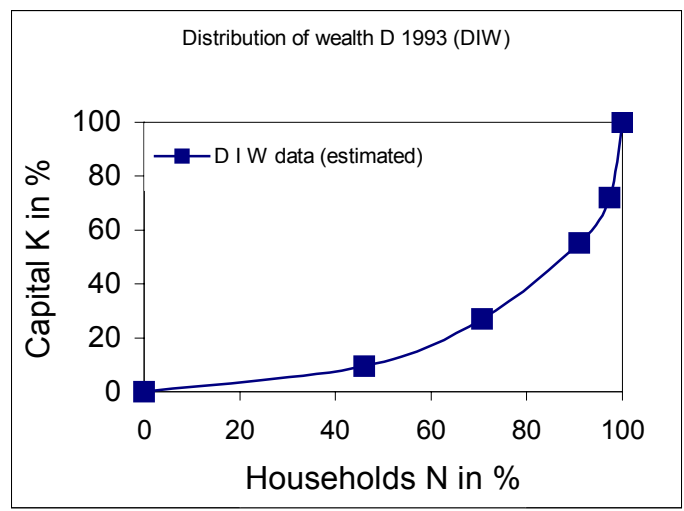

***Fig. 1.1*** *Lorenz distribution sum of capital K vs. sum of households N in Germany 1993, DIW (11).*

For a more detailed discussion the distribution N(k) of households and the amount of capital K(k) = k N(k) in each property class (k) will be compared and discussed for each model, separately.

The data of fig. 1.2 indicate that the values of $\alpha_j$ in the Cobb Douglas function seem to be constant, $\alpha_j = \alpha$, and eqs.(1.3) and (1.4) may be simplified as

$$N(k) = \alpha Y / (k/\lambda) \qquad (1.5),$$

$$K(k) = N k = \alpha \lambda Y = \text{constant} \qquad (1.6).$$

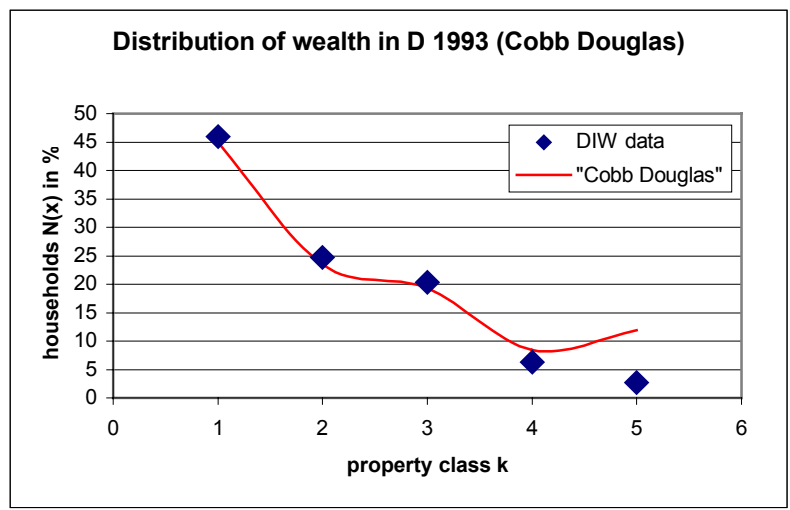

***Fig. 1.2*** *Distribution of households N(k) in property classes (k) in Germany 1993, data points according to DIW (11), curve according to Cobb Douglas, equ.(1,5).*

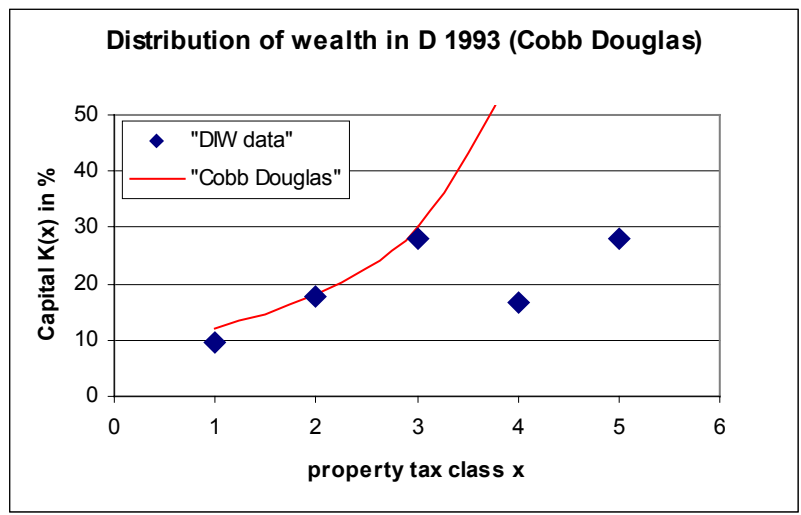

**Fig. 1.3** *Distribution of capital K(k) in property classes (k) in Germany 1993, data points according to DIW (11), curve according to Cobb Douglas, equ.(1,6).*

The data for N(k) in fig. 1.2 agree rather well with the Cobb Douglas function, the deviation from N(k) = 1/ k plot is due to the different width of the property classes (k) in table 1.

The data for K(k) in fig. 1.3 agree not so well with the Cobb Douglas function, the deviation from K(k) = constant is due to the different width of the property classes (k) in table 1.

However, a closer look reveals many deficiencies of eqs. (1.5) and (1.6) and the Cobb Douglas function:
1. The data for N (k) in fig. 1.2 agree only in a certain range with eq.(1.5) and do not fit very well large values of (k). For very low property classes (k) according to Cobb Douglas the number of households should go to infinity.
2. The data for large values of (k) K (k) in fig. 1.3 do not fit the Cobb Douglas model, eq.(1.6).
3. The total number of persons should be determined by the integral (from A = 0 to B = ∞)

$$\int N(k)\, dw \;=\; \alpha\, \lambda\, Y \int dw / w \;=\; \alpha\, \lambda\, Y\, [\ln B - \ln A] \qquad (1.7).$$

Both integral limits lead to infinite values for the total number of households or persons.

4. The total amount of capital should be determined by the integral (from A = 0 to B = ∞)

$$\int K(k) \, dk = \alpha \lambda Y \int dk = \alpha \lambda Y [B - A] \qquad (1.8).$$

Again the integral limit B leads to infinit values for the total capital.

These points indicate that the Cobb Douglas function (1.2) may not be the optimal function to reproduce the data in fig. 1.2 and 1.3. For this reason an alternative production function will now be discussed.

**2 Boltzmann distribution**
Striking rich, selling a car, finding the best marriage partner or hiring the best fitting people for a job is not only a question of economic reasoning, but also very much a matter of luck and chances. This leads to the idea, that the distribution of income may be a question of Lagrange probability with economic constraints like price or capital.

In stochastic systems with constraints the probability p has to be maximized and eq.(1.0) – according to Lagrange – has to be replaced by

$$\lambda \ln p + \Sigma N_j k_j \rightarrow \text{maximum!} \qquad (2.1)$$

The constraints are again given by $\Sigma N_j k_j$. We will now apply eq.(2.1) to the property distribution of table 1.

Distributing N households to k property classes is a question of combinatorial statistics,

$$p = N! / \Pi (N_j !) \qquad (2.2).$$

Using Sterling´s formula ( ln N! = N ln N - N) eq.(2.1) changes to

$$\lambda (N \ln N - \Sigma N_j \ln N_j) + \Sigma N_j k_j \rightarrow \text{maximum!} \qquad (2.3).$$

In this stochastic model the production function Y is given by the entropy function,

$$Y = (N \ln N - \Sigma N_j \ln N_j) = \ln (A \Pi N_j^{-N_j}) \qquad (2.4)$$

with $\Sigma N_j = N$. This function is similar to the Cobb Douglas function (1.2). However, the production function Y of the stochastic model is a logarithmic function and the elasticty constants $\alpha_j$ are replaced by $(-N_j)$.

At equilibrium (maximum) the derivative of equation (1.4) with respect to $N_j$ will again be zero, this leads to

$$\lambda \, \partial Y / \partial N_j - k_j = \alpha_j \, Y / N_j - k_j = 0$$

Solving this equation for $N_j$ we obtain the Boltzmann function

$$N_j(k_j) = A_0 \exp(-k_j/\lambda) \qquad (2.5).$$

The number of people in a property class ($k_j$) depends only on real property $k_j/\lambda$. The number of persons decreases exponentially with growing property. The total amount of property $K_j$ in the property class $k_j$ is given by

$$K_j(k_j) = N_j k_j = A_0 \, k_j \exp(-k_j/\lambda) \qquad (2.6).$$

We may test the Boltzmann function by comparing equations (2.5 and 2.6) to the property data in table 1.

The data of figs. 2.1 and 2.2 indicate that the eqs. (2.5) and (2.6) may be simplified

$$N(x) = A_0 \exp(-k/\lambda) \qquad (2.7),$$

$$K(w) = A_0 \, k \exp(-k/\lambda) \qquad (2.8).$$

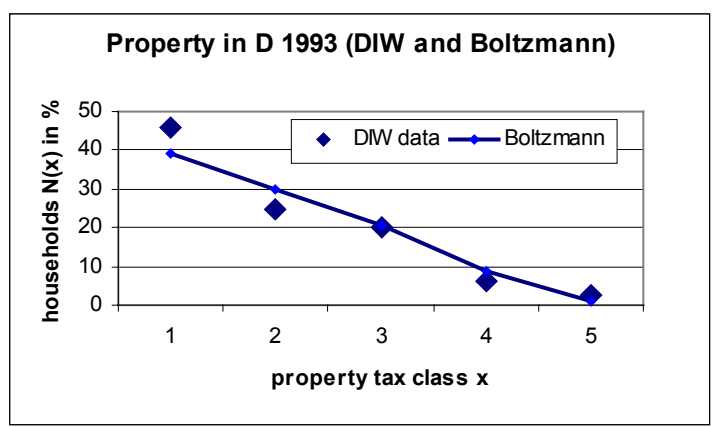

*Fig. 2.1* *Distribution of households N(k) in property classes (k) in Germany 1993, data points according to DIW (11), curve according to Boltzmann, equ.(2,7).*

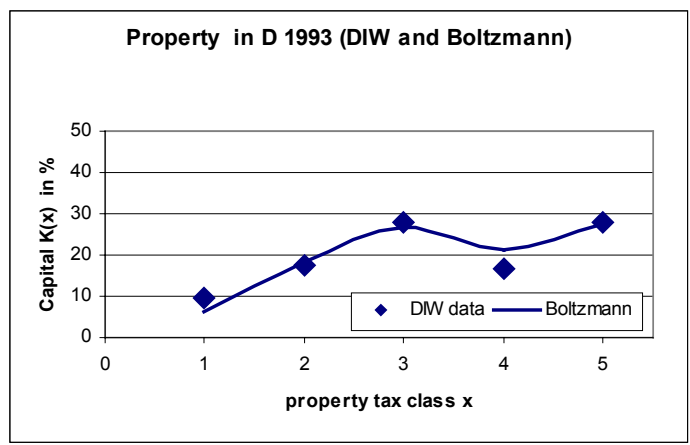

*Fig. 2.2* *Distribution of capital K(k) in property classes (k) in Germany 1993, data points according to DIW (11), curve according to Boltzmann, equ.(2,8).*

We will now take a closer look on these results:
1. The data for N (x) in fig. 2.1 agree rather well with eq.(2.7), but there are still small deviations.
2. The data for K (x) in fig. 2.2 also agree rather well with eq.(2.8), but there are again small deviations.
3. The total number of persons should be determined by the integral (from A = 0 to B = $\infty$)

$$\int N(x) \, dk = A_0 \lambda = N_0 = 19 \text{ Mill Hh} \qquad (2.9).$$

The integral leads to the correct number of total households and is used to calculate the constant $A_0$.

4. The total amount of property should be determined by the integral (from A = 0 to B = $\infty$)

$$\int K(k) \, dk = A_0 \lambda^2 = N_0 \lambda = K_0 = 190 \text{ Mrd. DM} \qquad (2.10).$$

The integral leads to the correct amount of total capital and determines the Lagrange parameter $\lambda$.

5. The Lagrange parameter $\lambda$ is determined by the ratio of total property and total number of households,

$$\lambda = K_0 / N_0 = 10.000 \text{ DM} \qquad (2.11)$$

and may be regarded as mean property of all households.

**Normal or Gaussian distribution in statistics without constraints**

Economic distributions are often fitted by a normal or Gaussian distributions. The normal distribution is a probability function that may be derived from the Lagrange principle of stochastic systems eq.(2.1), if the constraints are zero, $\Sigma N_j k_j = 0$:

$$\lambda \ln p + 0 \rightarrow \text{maximum!} \quad (2.12).$$

In systems without constraints the logarithm of probability and as well as probability p will tend to be at maximum. The probability is now given by

$$p(w) = \frac{1}{\sqrt{2\pi\overline{w}}} \exp\{-(w-\overline{w})^2/2\overline{w}\} \quad (2.13).$$

This probability may be applied to all problems without constraints. Economic systems require, that a specific distribution does not involve any costs, $\Sigma N_j k_j = 0$. Example: receiving more ore fewer phone calls generally does not involve additional costs, and we may expect normal distributions for phone calls. If the average number of phone calls is $\overline{w} = 30$ per day, we may wonder about the probability to receive today again w = 30 phone calls. The probability according to eq.(2.13) is about 7 %.

Economic distributions that involve money never should be Gaussian, but always Boltzmann.

## 3 Automobile market in Germany 1998

Table 2 shows the distribution of new cars in Germany in four classes,

| class j | cm³ | price class in DM | units in Mill. |
|---|---|---|---|
| | | | |
| 1 | < 15000 | 19.000 | 0,54 |
| 2 | > 1500 | 26.000 | 3,25 |
| 3 | > 2000 | 50.000 | 0,72 |
| 4 | > 2500 | 68.000 | 0,54 |

***Table. 2*** *Price classes ($k_j$) of automobiles produces in Germany (1998), (Stat. Bundesamt (12).*

Fig. 3.1 shows the number N(k) of new cars as a function of unit price (k).

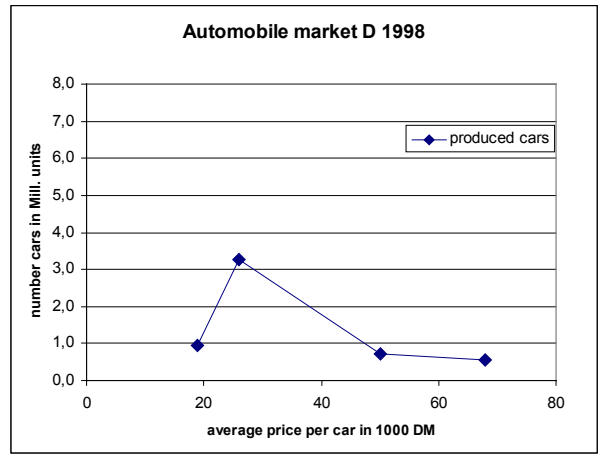

***Fig. 3.1*** *Distribution of new cars N(k) in price classes (k) in Germany 1998, data points according Stat. Bundesamt (12).*

The distribution of automobiles produced in Germany 1998 in fig. 3.1 may look like a normal distribution, but it should not be fitted by a normal distribution, buying cars always involves costs!
The car distribution can be fitted by a Cobb Douglas function, except for the lowest value at k = 20.000 DM. But since the Cobb Douglas function goes to infinity for low price values, this distribution cannot predict numbers for small values.

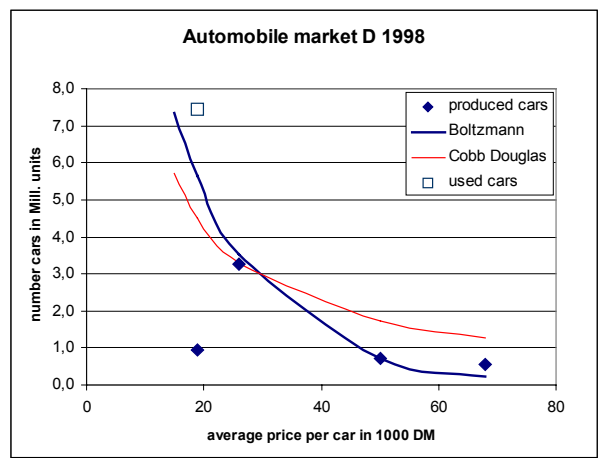

***Fig. 3.2*** *Distribution of new cars N(k) in price classes (k) in Germany 1998, data points according Stat. Bundesamt (12).The Boltzmann calculation predicts the correct number of used cars and fits better than Cobb Douglas function..*

The Boltzmann distribution can also be fitted to the data except for the lowest value (k = 20.000 DM), where a number of 7 Mill. units would be required. Apparently the new car market is not a complete market, and we have to add the used car market, as many buyers prefer cheaper used cars. Fig. 3.2 shows the complete market. The number of used cars sold in 1998 is nearly 7 Mill. units, as predicted by the Boltzmann distribution.

## 5 Marriage markets

The distribution of women and men seeking marriage may be calculated from the Lagrange principle, where probability is given by combinatorics.

$$M_j (P_j) = (M/W) \ a_j \ \exp(-P_j / \lambda) \quad (4.1).$$

The number of marriage proposals $M_j$ to a Lady grows with her attractiveness $a_j$ and is proportional to the ratio of all marriage seeking men M and women W. It will decrease with the expectations (bridal price $P_j$) of the Lady. On the other hand the chances $A_j$ of a young man to be accepted, grow with the ratio of men to women, with his attractiveness $a^*_j$ and the expectations (bridal price $P_j$) he is able to pay,

$$A_j (P_j) = (W/M) \ a^*_j \ \exp(P_j / \lambda) \quad (4.2).$$

Marriage is a question of supply and demand. For M>W we obtain fig. 4.1.

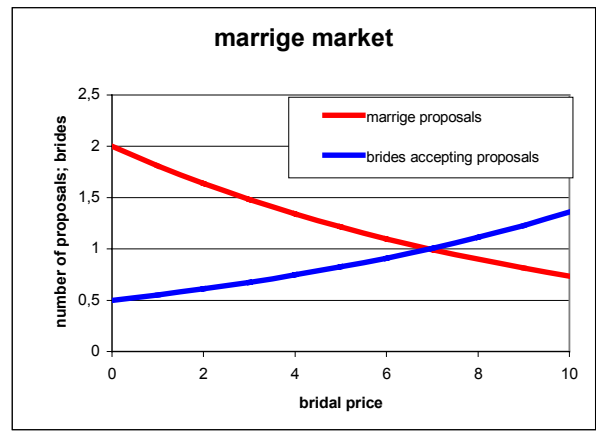

*Fig. 4.1* *Chances of marriage for more men than women. Women may have certain expectations (bridal price) for their future husbands. If the expectations are too high, the chances will drop.*

If there are more women than men, the men may have expectations (dowry), as shown in fig 4.2.

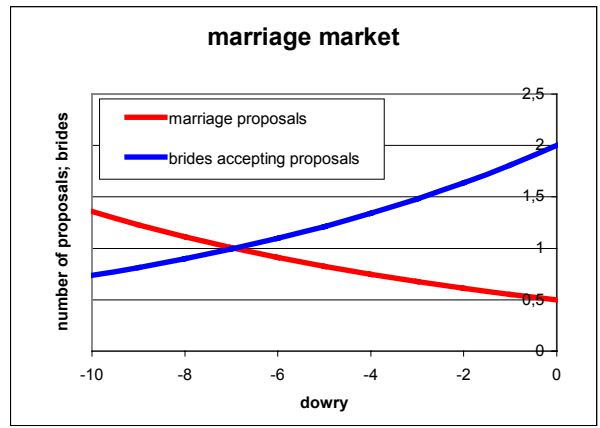

*Fig. 4.2* *Chances of marriage for more women than men. Men may have certain expectations (dowry) for their future bride.*

Fortunately, the number of women and men is nearly equal, so nobody has to pay very much. But especially attractive women can either get expect a bridal price from a normal husband or get a very attractive husband for nothing, fig. 4.3.

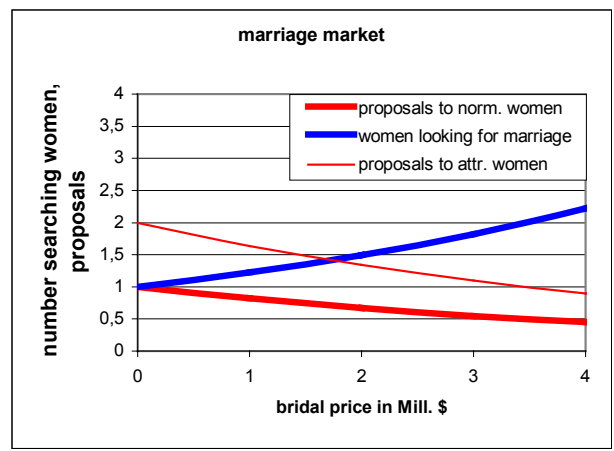

*Fig. 4.3* *Chances of marriage for equal number of women and men. Only very attractive women may have certain expectations (bridal price) for her future husband.*

The marriage market is ruled by general market laws of supply and demand. The mechanism stabilizes the different classes. Attractive (educated or rich) women will marry attractive (educated or rich) men. Poor people will marry poor people. Obviously, the market laws are governed by the Boltzmann distribution, by chance. Getting married, selling a car, striking rich is a matter of luck! And in contrast to the Cobb Douglas function the Boltzmann function can also lead to an equilibrium at a negative price!

**5  Job markets**

The distribution of people seeking jobs is a matter of supply and demand and may be calculated in the same way as for marriage. For the job market we obtain

$$J_j(w_j) = (J/P)\ a_j\ \exp(-w_j/\lambda) \qquad (5.1).$$

The number of job offers $J_j$ to a person grows with her attractiveness $a_j$ (education), the ratio of jobs to people and decreases with the wage expectations $w_j$. The chance to find an employee that accepts the job offer is

$$P_j(w_j) = (P/J)\ a^*_j\ \exp(w_j/\lambda) \qquad (5.2).$$

The number of people $P_j$ accepting an open position will grow with the attractiveness $a^*_j$ of the job, the ratio of people to jobs ($P/J$) and with the wage offered.

For many open jobs, $J > P$, a normal worker will be paid an equilibrium wage, only very good people will be able to earn more.

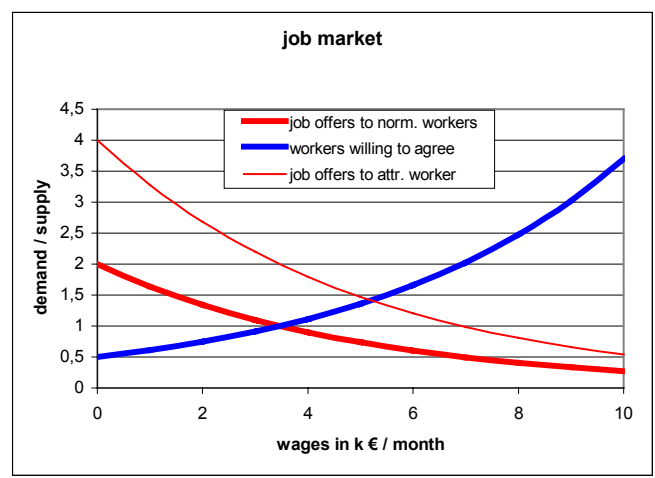

*Fig. 5.1* *Chances of finding a job at more jobs than people. A normal worker will be paid an equilibrium wage, only very good people may expect a better wage.*

At equal number of jobs and people the equilibrium wage is zero! However, as nobody can accept a zero pay job for a living, no average person can be hired. Only very good people will have a chance to get a good pay, the less than average people would have to pay the employer to work. Of course, this will not happen, so some jobs will be left open. This explains, why there are always open jobs even at high unemployment.

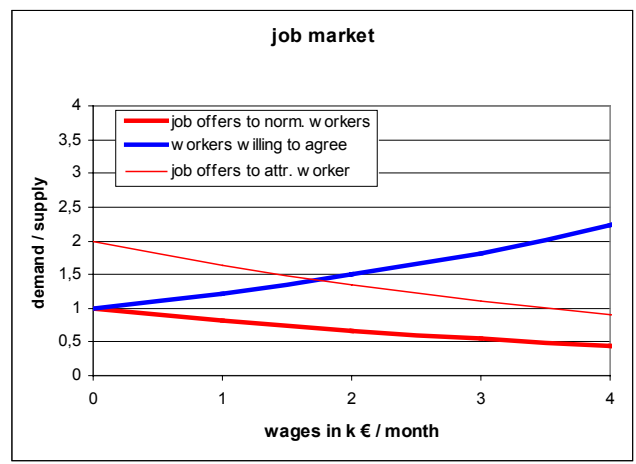

*Fig. 5.2* *Chances of finding a job for equal number of jobs and people. Only very good people may expect a wage for the future job. Normal and less attractive workers will not be hired, some jobs will stay open.*

The equilibrium wage may be calculated from the equilibrium of eqs.(5.1 and 5.2)

$$w / \lambda = \ln (J / P) + \tfrac{1}{2} \ln ( a_j / a^*_j ) \qquad (5.3).$$

The real wage depends on the logarithm $(J / P)$ of jobs to people and the logarithm of education of the worker and attractiveness of the job.

The actual distribution of jobs and wages may be expected to show a Boltzmann distribution according to eq.(5.1). However, jobs below a wage minimum $w_0$ have the attractiveness $a^* = 0$. Accordingly, the distribution of income will be given by

$$N (w) = a^*(w, w_0) \exp ( - w / \lambda ) \qquad (5.4).$$

The number of people earning a wage (w) will depend on the job attractiveness $a^*(w, w_0)$ and the Boltzmann function. Low wages will be very probable, high wages less probable. Figs. 5.3. and 5.4 show the wage distribution for service and production in the US in 1995 (12).

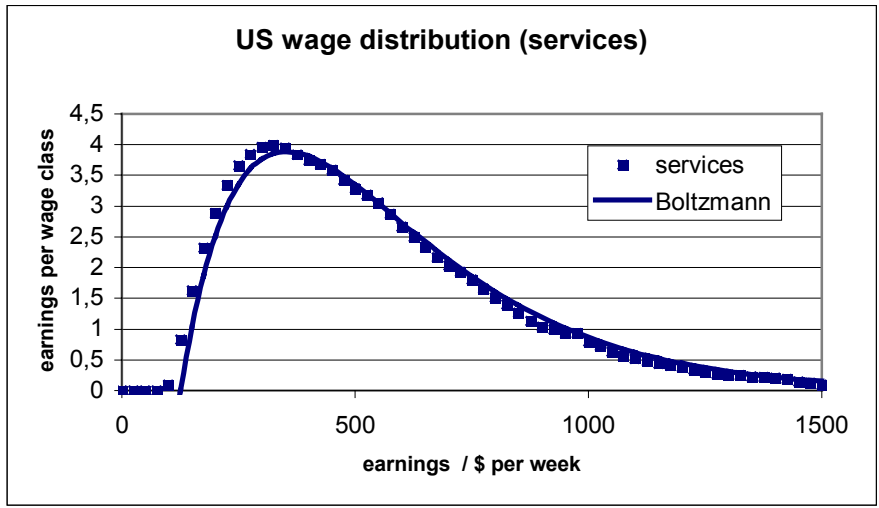

*Fig. 5.3* *Number of people in wage classes in services in the US (13). The data have been fitted by the Boltzmann distribution , eq.(5.4)*

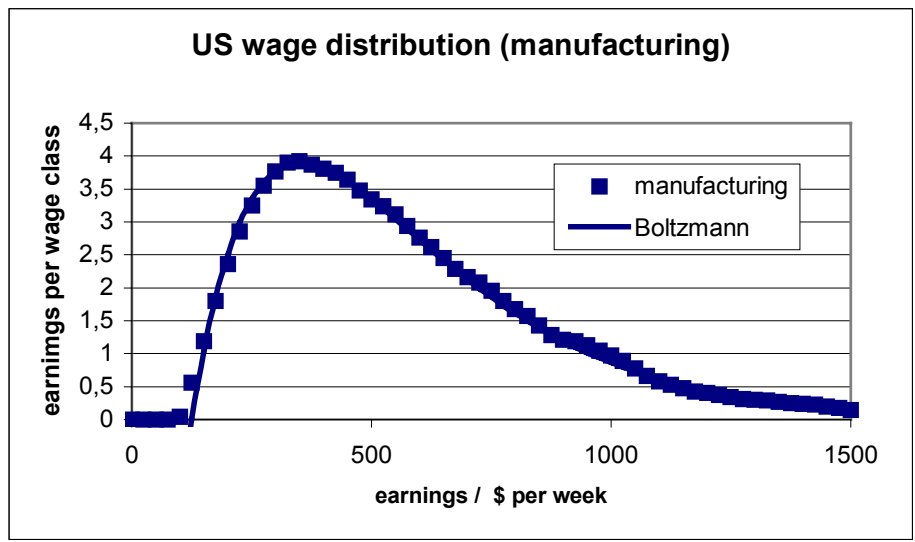

*Fig. 5.4* *Number of people in wage classes in manufacturing in the US (13) .The data have been fitted by the Boltzmann distribution , eq.(5.4).*

**Conclusion**

What has been gained by introducing the Boltzmann function to economics?

The Boltzmann function leads to a better understanding of economic interactions.

1. All economic systems may be regarded as stochastic systems with a probability that depends on costs and prices. Wealth and income are a matter of probability, as well as selling, buying, finding a job or getting married.

2. The Boltzmann function not only improves the Cobb Douglas distribution, but will also include other probability theories (game theory).

3. The Lagrange function links economic systems to social systems and natural systems like physics, chemistry, meteorology, which are also governed by the Lagrange principle. This may be one reason why many groups are now involved in econophysics and socio-economics.

4. Like in the other sciences the Lagrange principle may be regarded as the basis of economics, and most economic properties of the system may be calculated from the Lagrange principle! Only for time dependent properties a more general approach, e.g. the Fokker Planck equation (3) has to be applied. But for most problems the Lagrange principle will be sufficient.

**References**

(1) V. Pareto, Cours d'economie politique. Reprinted as a volume of Oeuvres Completes (Droz. Geneva, 1896-1965)

(2) N. Georgescu-Roegen, The entropy law and the economic process / Cambridge, Mass. Harvard Univ. Press, 1974.

(3) W. Weidlich, The use of statistical models in sociology, Collective Phenomena 1, (1972) 51

(4) D.K. Foley, A Statistical Equilibrium Theory of Markets, Journal of Economic Theory Vol. 62, No. 2, April 1994.

(5) J. Mimkes, Binary Alloys as a Model for Multicultural Society , J. Thermal Anal. 43 (1995) 521-537

(6) *M. Levy & S. Solomon, Power Laws are Logarithmic Boltzmann Laws, International Journal of Modern Physics C , Vol. 7, No. 4 (1996) 595;* J.

(7) Mimkes, Society as a many-particle System,  J. Thermal Analysis 60 (2000) 1055 - 1069

*(8)* M. Levy, H. Levy & S. Solomon, Microscopic Simulation of Financial Markets, Academic Press, New York, 2000.

(9)  P. Richmond & S. Solomon, Power Laws are Boltzmann Laws in Disguise, S. Solomon & P. Richmond, Stability of Pareto-Zif Law in Non-Stationary Economies, 2000

(10)   H. D. Kreft, Das Humanpotenzial, Verlag VWF, Berlin 2001

(11)   Report Deutsches Institut für Wirtschaftsforschung, DIW, Berlin 1993

(12)   Statistisches Jahrbuch 2000, Statistisches Bundesamt, Wiesbaden 2000

(13)   Cleveland Federal Reserve Bank; US Department of Labor; Bureau of Statistics. 1996